\documentstyle[preprint,prc,eqsecnum,aps,epsf]{revtex}

\tightenlines   %for single spacing in preprint

\begin{document}
\draft

\title{Three Body Bound State Calculations without Angular
Momentum Decomposition}

\author{ Ch.~Elster, W.~Schadow}
\address{
Institute of Nuclear and Particle Physics,  and
Department of Physics, \\ Ohio University, Athens, OH 45701}
    
\author{ A.~Nogga, W.~Gl\"ockle}
\address{
 Institute for Theoretical Physics II, Ruhr-University Bochum,
D-44780 Bochum, Germany.}

\vspace{10mm}

\date{\today}

\maketitle

\begin{abstract}   
The Faddeev equations for the three body bound state are
solved directly as three dimensional integral equation without
employing partial wave decomposition. The numerical stability 
of the algorithm is demonstrated. The three body binding energy
is calculated for Malfliet-Tjon type potentials and compared with 
results obtained from calculations based on partial wave decomposition.
The full three body wave function is calculated as function of the vector 
Jacobi momenta. It is shown that it satisfies the Schr\"odinger equation
with high accuracy. The properties of the full wave function are
displayed and compared to the ones of the corresponding wave functions
obtained as finite sum of partial wave components. The agreement
between the two approaches is essentially perfect in all respects.

\end{abstract}

\vspace{10mm}

%\pacs{21.45.+v,27.10.+h}

\pagebreak

%****************************************************************************
 \narrowtext 

%******************************************************************

\section{Introduction}

Three nucleon bound state calculations are traditionally carried out
by solving Faddeev equations in a partial wave truncated basis, which leads
to a set of a finite number of coupled equations in two variables
for the amplitude.  This is performed either in momentum
space \cite{pickle,statler,nbench}, in configuration space \cite{payne,schell0},
or in a hybrid fashion using both spaces \cite{wu}.
Though a few partial waves often provide qualitative 
insight, modern three nucleon bound state calculations
need 34 or more different isospin, spin
and orbital angular momentum combinations \cite{nbench}. 
It appears therefore natural 
to avoid a partial wave representation completely and work directly with 
vector variables.  This is common practice in bound state calculations of 
few nucleon systems based on variational \cite{wiringa} and Green's function 
Monte Carlo (GFMC) methods \cite{gfmc}. Those methods are normally
carried out in configuration space, and while providing accurate 
results for the binding energy, the GFMC method samples the 
wave function only stochastically. Up to now the Faddeev equations
have been solved applying vector variables only in configuration space
for pure Coulomb bound state problems, the $e^-e^-e^+$ and
$pp\mu^-$ systems \cite{kvin}.
There are other accurate techniques
for solving for the three body binding energy based on 
correlated Hyperspherical Harmonic variational techniques 
\cite{vivian,vivian2}
 and the  Gaussian 
basis coupled channel methods \cite{kamimura}, 
which, however, all work with partial wave decomposition.

Our aim is to work directly with vector variables in the Faddeev scheme
in momentum space. Under these circumstances 
the two-body $t$-matrices should also be 
determined directly as function of momentum vectors. This is not 
too difficult and has been recently demonstrated for two-body 
$t$-matrices based on Malfliet-Tjon type potentials \cite{3dt}. The
choice of momentum vectors as adequate variables is also suggested by
the NN force. Here the dependence on momentum vectors can be rather simple,
e.g., in the case of the widely used one-boson exchange force, whereas
the partial wave representation of this force leads to rather 
complicated expressions \cite{bonn}.

Instead of first summing up two-body interactions to infinite order 
into two-body $t$-matrices and obtain connected kernel equations, 
the Schr\"odinger equation for
the three nucleon (3N) bound state can be solved directly \cite{kalusch}.
In that work the Schr\"odinger equation was solved in momentum space for
identical particles without partial wave decomposition for Malfliet-Tjon 
type interactions. We do not want to follow this avenue but rather
solve the Faddeev equations in momentum space directly in a three 
dimensional form. Section II describes our solution of the Faddeev
equations as function of the momentum vectors and contrast this
new approach with the standard form given in a partial wave decomposition.
We work with spinless nucleons and thus formulate the Faddeev equations
for a system of three bosons.
In Section III we discuss
details of our algorithm for solving the Faddeev equations for the
three body bound state and demonstrate the stability of our numerical
approach as well as compare to results obtained from partial wave based
calculations. In Section IV we show properties of the Faddeev amplitude and
the full wave function and demonstrate that the results obtained from
the three dimensional approach and the ones coming from partial waves 
agree very well with each other. We compare momentum distributions and
expectation values calculated within the two schemes. In Section V
we demonstrate that the Schr\"odinger equation is fulfilled pointwise
by the full wave function generated in the three dimensional approach.
To the best of our knowledge this has never been demonstrated before with
this accuracy. 
We conclude in Section VI.

\section{Three-Body Bound State Equation}

The bound state of three identical particles which interact via pairwise
forces $V_i=V_{jk}$ ($ijk=123$ and cyclic permutations thereof) is given
by a Schr\"odinger equation which read in integral form
\begin{equation}
|\Psi \rangle = G_0 \sum_{i=1}^3 V_i \, | \Psi \rangle .\label{eq:2.1}
\end{equation}
Here the free propagator is given by $G_0 =(E-H_0)^{-1}$, and $H_0$
stands for the free Hamiltonian. Introducing Faddeev components
$|\Psi \rangle = \sum\limits_{i=1}^3 |\psi_i \rangle $ with
 $|\psi_i \rangle =G_0 \, V_i \, |\Psi \rangle$ leads to the 
 three coupled integral equations
\begin{equation}
| \psi_i \rangle = G_0 \, t_i \, \sum_{j\neq i} |\psi_j \rangle .
 \label{eq:2.2}
\end{equation}
The operator $t_i$ describes the two-body $t$-matrix in the
subsystem $jk$. If we consider identical particles (here bosons, since
we are omitting spin), the three nucleon wave function $|\Psi \rangle$ has to
be totally symmetric.  As a consequence, $|\psi_1 \rangle $, $|\psi_2
\rangle$, and $| \psi_3 \rangle $ are identical in their functional
form, only the particles are permuted. Thus it is sufficient to
consider only one component
\begin{equation}
|\psi \rangle =  G_0 \, t \,P\, |\psi \rangle  , \label{eq:2.3}
\end{equation}
where the arbitrarily chosen index $1$ is dropped. In this case the
permutation operator $P$ is given as $P= P_{12}P_{23}+P_{13}P_{23}$.
The complete three nucleon wave function is then given as
\begin{equation}
| \Psi \rangle  = (1 + P) \, | \psi \rangle . \label{eq:2.4}
\end{equation}

In order to solve Eq.~(\ref{eq:2.3}) in momentum space we introduce
standard Jacobi momenta
\begin{eqnarray}
{\bf p}_i &=& \textstyle{\frac{1}{2}}({\bf k}_j - {\bf k}_k)  \\ \nonumber
{\bf q}_i &=& \textstyle{\frac{2}{3}}({\bf k}_i - \textstyle{\frac{1}{2}}
({\bf k}_j+{\bf k}_k)),
         \label{eq:2.5}
\end{eqnarray}
where $ijk=123$ and cyclic permutations thereof. With the arbitrary
choice $i=1$ and omitting the index $1$ for clarity of presentation
Eq.~(\ref{eq:2.3}) reads
\begin{equation}
\langle {\bf p} {\bf q}|\psi\rangle = \frac{1}{ E- \frac{1}{m}p^2 -
 \frac{3}{4m} q^2} \, \langle {\bf p} {\bf q} |t P | \psi\rangle
\label{eq:2.6}
\end{equation}
with
\begin{equation}
\langle {\bf p} {\bf q}|t \, P | \psi\rangle = \int d^3 q'\, d^3p'\, d^3q'' \,
d^3p'' \; \langle {\bf p} {\bf q}|t|{\bf p}'{\bf q}'\rangle \,
\langle {\bf p}'{\bf q}'|P|{\bf p}''{\bf q}'' \rangle
\, \langle {\bf p}''{\bf q}''|\psi\rangle. \label{eq:2.7}
\end{equation}
The momentum states are normalized according to 
$\langle {\bf p}'{\bf q}'|{\bf p} {\bf q}\rangle = \delta^3 ({\bf p}'-
{\bf p}) \, \delta^3 ({\bf q}'-{\bf q})$. The Jacobi momenta in the
different two-body subsystems $(13)$ and $(12)$ are expressed through
those defined in the subsystem $(23)$ via
\begin{eqnarray}
{\bf q}_1 &=& -{\bf p}_2 - \textstyle{\frac{1}{2}} {\bf q}_2 
\nonumber \\ {\bf p}_1 &=&-\textstyle{\frac{1}{2}} {\bf p}_2 +
\textstyle{\frac{3}{4}} {\bf q}_2  \nonumber \\ {\bf q}_1 &=& {\bf p}_3
- \textstyle{\frac{1}{2}} {\bf q}_3  \nonumber \\ {\bf p}_1 &=& -
\textstyle{\frac{1}{2}} {\bf p}_3 - \textstyle{\frac{3}{4}} {\bf q}_3 .
\label{eq:2.8}
\end{eqnarray}
This allows to evaluate the permutation operator given in
Eq.~(\ref{eq:2.7}) as 
\begin{eqnarray}
\langle {\bf p}'{\bf q}'|P|{\bf p}'' {\bf q}''\rangle &=&
  \langle {\bf p}'{\bf q}'|{\bf p}'' {\bf q}''\rangle_2 +
  \langle {\bf p}'{\bf q}'|{\bf p}'' {\bf q}''\rangle_3  \nonumber \\
 &=& \delta^3 ({\bf p}'+\textstyle{\frac{1}{2}}{\bf q}'+{\bf q}'') \,
     \delta^3 ({\bf p}''-{\bf q}'-\textstyle{\frac{1}{2}}{\bf q}'')
 \nonumber \\
 & &  + \,
     \delta^3 ({\bf p}'-\textstyle{\frac{1}{2}}{\bf q}'-{\bf q}'') \,
     \delta^3 ({\bf p}''+{\bf q}'+\textstyle{\frac{1}{2}}{\bf q}''),
 \label{eq:2.9}
\end{eqnarray} 
where the indices $2$ and $3$ indicate the corresponding subsystem.
Inserting this relation into Eq.~(\ref{eq:2.7}) and then into
Eq.~(\ref{eq:2.6}) leads to the expression for the Faddeev amplitude
which serves as starting point for our numerical calculations,
\begin{equation}
\langle {\bf p}{\bf q}|\psi\rangle = \frac{1}{ E- \frac{1}{m} p^2 -
 \frac{3}{4m} q^2} \, \int d^3q' \; t_{\rm s}({\bf p},\textstyle{\frac{1}{2}}
 {\bf q}+{\bf q}', E-\textstyle{\frac{3}{4m}} q^2) \, \langle {\bf
 q}+\textstyle{\frac{1}{2}}{\bf q}',{\bf q}'|\psi\rangle.  \label{eq:2.10}
\end{equation}
Here $t_{\rm s}({\bf p},{\bf q},E)$ is the symmetrized two-nucleon $t$-matrix,
\begin{equation}
t_{\rm s}({\bf p},{\bf q},E)=t({\bf p},{\bf q},E)+t(-{\bf p},{\bf q},E) .
\end{equation}
\noindent
We would like to mention
that the so obtained Faddeev amplitude fulfills the 
symmetry relation $\psi({\bf p}{\bf q}) = \psi(-{\bf p}{\bf q})$, as
can be seen from Eq.~(\ref{eq:2.10}).

In Eq.~(\ref{eq:2.10}) the Faddeev amplitude $\psi({\bf p}{\bf q})$ is
given as function of vector Jacobi momenta and obtained as solution of a
three dimensional integral equation. For the ground state 
$\psi({\bf p}{\bf q})$ is a scalar and thus depends only on the
magnitudes of ${\bf p}$ and ${\bf q}$ and the angle between those
vectors.  In order to solve this equation
directly without inserting partial wave projections, we have to define
a coordinate system. We choose the vector ${\bf q}$ parallel to the
z-axis and express the remaining vectors ${\bf p}$ and ${\bf q}'$ with
respect to ${\bf q}$. As variables we have the magnitudes of the vectors
as well as the following angle relations,
\begin{eqnarray}
x &\equiv& {\hat {\bf p}}\cdot {\hat {\bf q}} = \cos \theta \nonumber \\
 x'&\equiv& {\hat {\bf q}'}\cdot {\hat {\bf q}} =\cos \theta' \nonumber \\
 y &\equiv& {\hat {\bf p}}\cdot {\hat {\bf q}'} = \cos
\gamma , \label{eq:2.11}
\end{eqnarray}
where
\begin{equation}
\cos \gamma = \cos \theta \cos \theta' + \sin \theta \sin \theta'
 \cos (\varphi-\varphi') = xx' +\sqrt{1-x^2} \; \sqrt{1-x'^2} \,\cos
 \varphi'.
\label{eq:2.12}
\end{equation}
To arrive at the last relation we took advantage of the freedom of
choice for one azimuthal angle and set $\varphi=0$. We have this freedom
due to the $\varphi'$ integration over the full $2\pi$ interval.

\noindent
With these variables Eq.~(\ref{eq:2.10}) can be explicitly written as
\begin{eqnarray}
\psi(p,q,x)&=& \frac{1}{ E- \frac{1}{m}p^2 -\frac{3}{4m} q^2}
 \int\limits_0^{\infty} dq' q'^2 \int\limits_{-1}^1 dx'
 \int\limits_0^{2\pi} d\varphi' \nonumber \\ & & \times \, t_{\rm s}
 \left(p,\sqrt{\textstyle{\frac{1}{4}} q^2+q'^2+qq'x'},
 \frac{\frac{1}{2}qx+q'y} {|\frac{1}{2}{\bf q}+{\bf q}'|};
 E-\textstyle{\frac{3}{4m}} q^2 \right) \nonumber \\ & &  \times \, \psi
 \left(\sqrt{q^2+\textstyle{\frac{1}{4}}q'^2+qq'x'}, q',
 \frac{qx'+\frac{1}{2}q'}{|{\bf q}+\frac{1}{2}{\bf q}'|} \right),
 \label{eq:2.13}
\end{eqnarray}

\noindent
where $|\frac{1}{2}{\bf q}+{\bf q}'|=
\sqrt{\frac{1}{2}q^2+q'^2+qq'x'}$ and $|{\bf q}+\frac{1}{2}{\bf q}'|
= \sqrt{q^2+\frac{1}{4}q'^2+qq'x'}$.

\noindent
The above equation, Eq.~(\ref{eq:2.13}), is the starting point for
our numerical algorithms, and the details will be described in the next
Section.

In a standard partial wave representation \cite{wbook} Eq.~(\ref{eq:2.13})
is replaced by a set of coupled two-dimensional integral equations
\begin{eqnarray}
\label{partialfad}
\psi_l(p,q) &=& \frac{1}{ E- \frac{1}{m}p^2 -\frac{3}{4m} q^2} \nonumber \\
& & \times \sum_{l'} \int\limits_0^{\infty} dq'  q'^2\int\limits_{-1}^1 dx' \;
  \frac{t_l(p,\pi_1,E-\textstyle{\frac{3}{4m}} q^2)}{\pi_1^l} \;
  G_{ll'}(q,q',x') \; \frac{\psi_l'(\pi_2,q')}{\pi_2^{l'}}, \label{eq:2.13a}
\end{eqnarray}
where
\begin{eqnarray}
\pi_1 &=& \sqrt{q'^2 +\textstyle{\frac{1}{4}}q^2 +qq'x'} \nonumber \\
\pi_2&=& \sqrt{q^2 +\textstyle{\frac{1}{4}}q'^2 +qq'x'}. \label{eq:2.13b}
\end{eqnarray}
Here $G_{ll'}(q,q',x')$ together with the shifted arguments 
$\pi_1$ and $\pi_2$ are the partial wave  representation of
the permutation operator.
The explicit form of  $G_{ll'}(q,q',x')$ can 
be found in Ref.~\cite{wbook}. However, the expression given
in this reference  is
more complicated due to spin and isospin variables. For the convenience
of the reader we give $G_{ll'}(q,q',x')$ defined for general total
orbital angular momentum in Appendix A explicitly.
The quantities $t_l$ are the partial wave projected two-body $t$-matrices.
Due to the symmetry requirement of the Faddeev amplitude,
$l$ runs over even values only. The
infinite set of coupled integral equations given in Eq.~(\ref{eq:2.13a})
is truncated in actual calculations at a sufficiently high value of $l$.
The full Faddeev amplitude $\psi(p,q,x)$ reads then
\begin{equation}
\psi(p,q,x) = \sum_l \frac{\sqrt{2l+1}}{4\pi} P_l(x) \, \psi_l (p,q).
  \label{eq:2.13c}
 \end{equation}

\vspace{3mm}
From the Faddeev amplitude $\psi({\bf p},{\bf q})$  the three
nucleon wave function is obtained by adding the components defined
in  the different subsystems as given in Eq.~(\ref{eq:2.4}). After
evaluating the permutation operator $P$, the wave function is given
as
\begin{equation}
\Psi({\bf p},{\bf q}) = \psi({\bf p},{\bf q})
+\psi(-\textstyle{\frac{3}{4}}{\bf q}-\textstyle{\frac{1}{2}}{\bf
p},{\bf p}-\textstyle{\frac{1}{2}}{\bf q})
+\psi(\textstyle{\frac{3}{4}}{\bf q}-\textstyle{\frac{1}{2}}{\bf
p},-{\bf p}-\textstyle{\frac{1}{2}}{\bf q}).
\label{eq:2.14}
\end{equation}
Already here we see that $\Psi({\bf p},{\bf q})=\Psi(-{\bf p},{\bf q})$,
provided the Faddeev components fulfill this symmetry. Again 
the momentum ${\bf q}$ is chosen in the direction of the z-axis 
and after some algebra the
explicit expression for the three nucleon wave function reads
\begin{eqnarray}
\Psi(p,q,x)&=& \psi(p,q,x) \nonumber \\ && + \, \psi
\left(\textstyle{\frac{1}{2}}\sqrt{\textstyle{\frac{9}{4}}q^2+p^2+3pqx},
\sqrt{p^2+\textstyle{\frac{1}{4}}q^2-pqx}, \displaystyle{\frac{
\frac{3}{8}q^2-\frac{1}{2}p^2-\frac{1}{2}pqx} {|-\frac{3}{4}{\bf
q}-\frac{1}{2}{\bf p}||{\bf p}- \frac{1}{2}{\bf q}|}}\right) \nonumber
\\ && + \, \psi
\left(\textstyle{\frac{1}{2}}\sqrt{\textstyle{\frac{9}{4}}q^2+p^2-3pqx},
\sqrt{p^2+\textstyle{\frac{1}{4}}q^2+pqx},\displaystyle{
\frac{-\frac{3}{8}q^2+\frac{1}{2}p^2-\frac{1}{2}pqx} {|
\frac{3}{4}{\bf q}-\frac{1}{2}{\bf p}||-{\bf p}- \frac{1}{2}{\bf
q}|}}\right), \label{eq:2.15}
\end{eqnarray}
where the magnitudes in the denominators of the angle variables are
given by
\begin{eqnarray}
\left|\textstyle{\frac{3}{4}}{\bf q} \pm \textstyle{\frac{1}{2}}{\bf
 p}\right| &=& \textstyle{\frac{1}{2}}
 \sqrt{\textstyle{\frac{9}{4}}q^2 +p^2 \pm 3pqx}  \nonumber \\
 \left|{\bf p} \pm \textstyle{\frac{1}{2}}{\bf q}\right| 
  &=& \sqrt{ p^2 + \textstyle{\frac{1}{4}}q^2 \pm pqx}.  \label{eq:2.16}
\end{eqnarray}

\noindent
The wave function is normalized according to
\begin{equation}
\int d^3 p \; d^3 q \; \Psi^2 ({\bf p},{\bf q})=1,  \label{eq:2.17}
\end{equation}
which reads explicitly
\begin{equation}
8 \pi^2 \int\limits_0^{\infty} p^2 dp \int\limits_0^{\infty} q^2 dq
 \int\limits_{-1}^1 dx \; \Psi^2(p,q,x) =1. \label{eq:2.18}
\end{equation}
The properties of the three nucleon wave function will be studied
in Section IV.

In a partial wave representation Eq.~(2.18) takes the form
\begin{equation}
\Psi(p,q,x) = \sum_l       \frac{\sqrt{2l+1}}{4\pi} P_l(x) \,
   \Psi_l(p,q), \label{eq:2.19}
   \end{equation}
where
\begin{equation}
\label{partialwave}
\Psi_l(p,q) = \psi_l(p,q) + \sum_{l'} \int\limits_{-1}^1 dx \,
 \tilde G_{ll'} (p,q,x)\, \psi_{l'}(\tilde \pi_1,\tilde \pi_2)
   \label{eq:2.20}
 \end{equation}
and
\begin{eqnarray}
\tilde \pi_1 &=& \sqrt{\textstyle{\frac{1}{4}}p^2+
 \textstyle{\frac{9}{16}}q^2 +\textstyle{\frac{3}{4}}qpx}  \nonumber \\
\tilde \pi_2&=& \sqrt{p^2 +\textstyle{\frac{1}{4}}q^2 -qpx}. 
\label{eq:2.21}
\end{eqnarray}
The quantity $\tilde G_{ll'} (p,q,x)$, which results from
applying the permutation operator, can be found in Ref.~\cite{hub2},
but should be considered without spin and isospin factors. 
The resulting expression for the here considered 
 bosonic case is given in Appendix A.

\section{Calculation of the Three-Body Binding Energy}

For our model calculations we use Yukawa interactions of 
Malfliet-Tjon \cite{MT} type, 
\begin{equation}
V({\bf p'},{\bf p})= \frac{1}{2\pi^2}\left( \frac{V_{\rm R}}{({\bf
      p'}- {\bf p})^2 + \mu_{\rm R}^2} - \frac{V_{\rm A}}{({\bf
      p'}-{\bf p})^2 + \mu_{\rm A}^2} \right).  \label{eq:3.1}
\end{equation}
We study two different types of pairwise forces, a purely attractive
Yukawa interaction and a superposition of a short-ranged repulsive and a
long-ranged attractive Yukawa interaction. In order to be able to
compare our calculations with results obtained by other methods we use
the Malfliet-Tjon potentials MT-IV and MT-V \cite{MT}. However, we use
slightly different parameters as given in Ref.~\cite{MT} to compare with
Refs.~\cite{schell0,schell2} as well as
 \cite{payne} and give our parameters in
Table~I. The calculated
values for the deuteron binding energy are $E_{\rm d}=-2.2087$~MeV 
for the MT-IV potential and $E_{\rm d}=-0.3500$~MeV for the MT-V potential,
respectively.
We also need to point out  that we calculate the potentials as
functions of vector momenta and thus define the interaction 
as a truly local force acting in all partial waves.

With this interaction we first solve  the Lippmann-Schwinger equation for
the fully-off-shell two-nucleon $t$-matrix directly as function of the
vector variables as described in detail in Ref.~\cite{3dt}.
The so obtained  $t$-matrix is then symmetrized to get
$t_{\rm s}( p', p, x;E-\frac{3}{4m}q^2)$. We would like to point out that
after having solved the Lippmann-Schwinger equation on Gaussian grids
for $p$, $p'$, and $x$, we solve the integral equation again to obtain the
$t$-matrix at points $x=\pm 1$. Thus, when iterating
Eq.~(\ref{eq:2.13}), we do not have to 
extrapolate numerically to  angle points $x$ of 
$t_{\rm s}( p', p, x;E-\frac{3}{4m}q^2)$, which can
very well be located outside the upper or 
lower boundary of the Gaussian grid of the $t$-matrix.

The eigenvalue equation, Eq.~(\ref{eq:2.13}), is solved by iteration.
Schematically this can be written as 
\begin{equation}
\lambda \, \psi = {\cal K}(E) \, \psi,  \label{eq:3.2}
\end{equation}
and we search for $E$ such that $\lambda = 1$. The
functional behavior of ${\cal K}(E)$ is determined by the two-body
$t$-matrix, and $\lambda =1$ is always the largest
positive eigenvalue regardless of the two-nucleon potential being employed.
For this reason the  simple iteration starting with an
essentially  arbitrary vector is sufficient. We start with a
vector $\psi(p,q,x) \sim 1/((1+p^2)(1+q^2))$  and stop the iteration when 
Eq.~(\ref{eq:3.2}) is fulfilled with a relative accuracy of $10^{-10}$
at each point $(p,q,x)$.

In order to solve the eigenvalue equation, Eq.~(\ref{eq:2.13}), 
for the Faddeev component $\psi(p,q,x)$  we use Gaussian grid points
in $p$, $q$, and $x$, though only the $q$-variable is a true integration
variable. Typical grid sizes are
 $97 \times 97 \times 42$ to obtain an accuracy in the
binding energy of 4 significant figures. This grid contains 
$x$ points for $x=\pm1$ as well as the points $p=0$ and $q=0$. 
%These
%additional points are not contained in a Gaussian grid and have to
%be solved for separately.
 The iteration of Eq.~(\ref{eq:2.13}) requires
a two-dimensional interpolation on  $\psi(p,q,x)$ in the
variables $p$ and $x$.  By including
the additional grid points, we avoid the extrapolation outside the 
Gaussian grid.
 The $q'$-integration in 
Eq.~(\ref{eq:2.13}) is cut off at a value of $q_{\rm max} = 20$~fm$^{-1}$.
The integration interval is divided into two parts, 
$(0,q_0)\cup (q_0,q_{\rm max})$, in which we use Gaussian quadrature
with NQ1 and NQ2 points respectively. The value for
$q_0$ is chosen to be 10~fm$^{-1}$. Typical values for NQ1 and
NQ2 are 64 and 32 to obtain the above mentioned accuracy. For the
distribution of quadrature points we use the maps given in
Ref.~\cite{langank}. The $x'$ integration requires typically at least 
32 integration points, while for the $\varphi'$ integration 20 points
are already sufficient. The $\varphi'$ integration acts only on
$t_{\rm s}$ and can thus be carried out before starting to solve
Eq.~(\ref{eq:2.13}). The $p$ variable is also defined in an
interval $(0,p_0)\cup (p_0,p_{\rm max})$, where $p_0$ is chosen to be
9~fm$^{-1}$ and $p_{\rm max}=60$~fm$^{-1}$. The two intervals contain
NP1 and NP2 points, and we choose NP1 = NQ1 and NP2 = NQ2 respectively.

The momentum dependencies given in Eq.~(\ref{eq:2.13}) suggest that we
solve the two-body $t$-matrix $t_{\rm s}(p,p',x;\varepsilon)$ on the momentum
grid $p$ and $p'$ for the energies $\varepsilon = E - \frac{3}{4m}
q^2$ dictated by the same $q$-grid.  It turns out that it is
sufficient to choose for the variable $x$ the same grid which is used
for solving the two-body Lippmann-Schwinger equation for
$t_{\rm s}(p,p',x;\varepsilon)$.  In order to obtain the second momentum and
the angle for $t_{\rm s}(p,p',x;\varepsilon)$ required in the integration of
Eq.~(\ref{eq:2.13}), we have to carry out two-dimensional
interpolations. We use the Cubic Hermitean splines of Ref.~\cite{hub}.
The functional form of those splines is described in detail in
Appendix~B of this reference and shall not be repeated here. We find
these splines superior to standard B-splines \cite{deboer} in
capturing the peak structure of the two-body $t$-matrix, which occurs
for off-shell momenta $p \simeq p'$.  In Fig.~1 the
off-shell structure of the symmetrized $t$-matrix $t_{\rm s}(p, p',
x;\varepsilon)$ for a center-of-mass (c.m.) energy
$\varepsilon=-50$~MeV and an off-shell momentum
$p'=0.75$~fm$^{-1}$ is displayed. 
The two steep dips at $x=\pm 1$ dominate the
off-shell structure and are present for all off-shell momenta $p
\simeq p'$.  An additional advantage of the Cubic Hermitean splines is
their computational speed, which is an important factor, since the
integral in Eq.~(\ref{eq:2.13}) requires a very large number of
interpolations. Finally, to obtain the Faddeev component entering the
integration we also need a two-dimensional interpolation and use the
same as for the $t$-matrix.

In Table~II we show the convergence of the three
nucleon binding energy as function of the number of grid points 
for the Malfliet-Tjon potentials MT-V and MT-IV. 
We use the potential parameters as given in Table~I. 
In Table~III  the convergence of the energy eigenvalues for
the same potentials resulting from the solution of the partial
wave Faddeev equation for an optimized set of $p$ and $q$ grid points
as function of the angular momentum $l$ is given.

As demonstrated in Table~II, the calculation of the three nucleon
binding energy using the Malfliet-Tjon potential V converges to a
value of $E_{\rm t}=-7.7365$~MeV. The solution of the Faddeev
equation in partial waves      converges to $E_{\rm t}=-7.7366$~MeV. Here
convergence is reached for $l=12$. Both of the here calculated values
for the potential MT-V are in excellent agreement with the value
$E_{\rm t}=-7.7366$~MeV given in Ref.~\cite{schell2}.  The agreement
with the slightly older calculation of Ref.~\cite{payne}, which uses 6
partial waves for the potential and yielding $E_{\rm t}=-7.736$~MeV
is also excellent. The agreement with the value 
$E_{\rm t}=-7.7365$~MeV obtained from a the paired potential 
basis calculation \cite{vivian} gives the same excellent agreement.

For the purely attractive potential MT-IV our calculations of the
binding energy converge to $E_{\rm t}=-25.050$~MeV as shown in
Table~II.  The solution based on partial wave decomposition converges
to $E_{\rm t}=-25.057$~MeV for $l = 12$.  Both values are also in
very good agreement with the value $E_{\rm t}=-25.05$~MeV quoted in
Ref.~\cite{kalusch}. As we can see from these comparisons to
calculations of the three nucleon binding energy based on the solution
of the Faddeev equations in partial waves, our results provide the
same accuracy while the numerical procedures are actually easier to
implement.  In the three dimensional case there is only one single
equation to be solved, whereas in the partial wave case one has a set
of coupled equations with a kernel containing relatively complicated
geometrical expressions $G_{ll'}(p,p',x)$ and $t_l(p,p',x)$ matrices
to larger values of $l$.  The latter ones are driven by the partial
wave projected potential matrix elements $V_l(p,p')$, which require
for large $l$-values and larger $p$-values (80-100 fm$^{-1}$) great
care to be generated reliably.  In contrast, the $t$-matrices are generated
more easily as functions of vector momenta from the original
potential.  Once an accurate and fast two dimensional interpolation
scheme is established, the implementation of Eq.~(\ref{eq:2.13}) is
rather simple.

\section{The Three Nucleon Wave Function}
\subsection{Properties}

When solving Eq.~(\ref{eq:2.13}) for the bound state, we 
obtain  the Faddeev component $\psi(p,q,x)$, and the symmetry
relation for the Faddeev component reads explicitly $\psi(p,q,x)=
\psi(p,q,-x)$. Note that in Eq.~(\ref{eq:2.13}) this symmetry is
only fulfilled because of the $\varphi$-integration over the complete 
2$\pi$ interval.
We verified that our numerical solution satisfies
this symmetry with high accuracy. In fact, this symmetry property can
be implemented in the iteration of Eq.~(\ref{eq:2.13}) 
to cut down the size of the
field $\psi(p,q,x)$ and thus save time and memory when computing the
integral. We would like to remark that for the initial calculations we
did not take advantage of this symmetry property in order to use
the numerical verification as a test for the accuracy of the integration
and interpolation.

In Fig.~2 we display magnitude of the Faddeev component at a fixed
angle, \mbox{$\psi(p,q,x=1)$} obtained from a calculation based on
the MT-V potential.  The norm is chosen such that 
$\langle\Psi|\Psi\rangle =
3\,\langle \psi| \Psi \rangle=1$, where $\Psi$ is the full 3N wave
function as given in Eq.~(2.18).  The major contributions to
$\psi(p,q,x)$ arise from momenta $p$ and $q$ less than
1.5~fm$^{-1}$. It also turns out that the dependence of $\psi(p,q,x)$
on the angle between the Jacobi momenta ${\bf p}$ and ${\bf q}$ is so
weak that it can only  be detected on a logarithmic
scale but not when comparing linear plots at
different values of x. This is understandable, since the admixture of
the $l>0$ partial waves into the Faddeev amplitude is less than 0.01\%.
To show this we choose the normalization differently from above as
\begin{equation}
1 = \int\limits_{-1}^1 dx \int\limits_0^{\infty}  dp \, p^2
      \int\limits_0^{\infty} dq \, q^2 \, \psi^2(p,q,x) 
\end{equation}
or, according to Eq.~(\ref{eq:2.13c})
\begin{eqnarray}
1  &=& \sum_l \int\limits_0^{\infty}  dp \, p^2 
     \int\limits_0^{\infty} dq \, q^2 \, \psi^2_l(p,q) \nonumber \\
 &\equiv & \sum_l F_l. \label{eq:4.1}
\end{eqnarray}

\noindent
Similar we define for the full wave function

\begin{eqnarray}
1  &=& \sum_l \int\limits_0^{\infty}  dp \, p^2 
     \int\limits_0^{\infty} dq \, q^2 \, \Psi^2_l(p,q) \nonumber \\
 &\equiv & \sum_l W_l. 
\end{eqnarray}

\noindent
The partial wave contributions $F_l$ and $W_l$ are given in Table~IV for
the different $l$ values. 

The logarithmic presentation in Fig.~2 allows to see not only the smooth
bell-like decrease of the Faddeev amplitude for $p$ and $q$ less
than 1.5~fm$^{-1}$ but also the deep valleys caused by the node lines
of the amplitude. A more detailed inspection of the Faddeev amplitude
is presented in Fig.~3. Here the partial wave amplitudes $\psi_l(p,q)$
are projected out numerically from the full amplitude $\psi(p,q,x)$
and compared to the amplitudes $\psi_l(p,q)$ directly determined from
the coupled set of equations given in Eq.~(\ref{eq:2.13a}). We show
contour plots for the $l$ values 0, 2, and 4. 
One can see that for the higher $l$ values the maximum of the
partial wave amplitudes moves to larger $p$ and $q$ values. It is also
apparent from the figure that the graphs representing  the 
three dimensional calculation and those for ones based on
partial waves  are nearly everywhere indistinguishable. This 
underlines the accuracy and consistency in the numerical realization of
both schemes. Only for $l=4$ and very small $q$-values deviations 
occur. There
the contour lines have to be almost parallel to the $p$-axis, indicating
the proper threshold behavior for very small $q$.
This is not the case in the 3D approach. However, the value of the
amplitude is smaller by 6 orders of magnitude compared to the 
value of the the $l=0$ amplitude at these values. In both cases
we hit numerical inaccuracies here for the number of grid points
chosen.

The evaluation of the 3N wave function according to Eq.~(2.18)
requires interpolations in three dimensions. We would like to 
reiterate here, that for the solution of the Faddeev equation,
Eq.~(\ref{eq:2.13}),  only  two-dimensional interpolations
are required. To carry out the 3D interpolation we use standard 
B-splines as given in \cite{deboer}.   In case of the wave function
B-splines turned out to be slightly more accurate on the given grids 
compared to the Cubic Hermite splines. Since the wave function is
calculated only once, the computational speed is not an issue 
here.

The absolute value of the
total wave function $\Psi(p,q,x)$ calculated
from the MT-V potential for a fixed angle $x=1$ is displayed in Fig.~4.
This wave function has
a smooth shape and is significant in size only for $p$ and $q$
values smaller than 1.5~fm$^{-1}$. 
It is interesting to see that the shape of the wave function is quite
similar to that of the Faddeev  amplitude of Fig.~2.  The function
values are of course different.

In Fig.~5  the partial wave projected amplitudes $\Psi_l (p,q)$ 
are displayed as contour plots  for $l = 0$, 2, and 4.
The left column contains the $\Psi_l (p,q)$ obtained from the
solution of Eq.~(\ref{eq:2.15}), whereas the right column contains
$\Psi_l (p,q)$ calculated directly in partial waves from Eq.~(2.24).
Again, both
plots are nearly everywhere indistinguishable, underlining the accuracy and
consistency in the numerical realization of both schemes.
For $l=4$  the contour plot for 
$\Psi_4 (p,q)$ calculated in partial waves show small irregularities
for $p$ values larger than 5~fm$^{-1}$ and 
 $q$ close to 0. These irregularities are  missing in the
projections obtained from the full 3D solution, but again as pointed
out above the 3D solution is not correct there either.

In Fig.~6 we show the wave function $\Psi(p,q,x)$ calculated from the
MT-IV potential. A comparison
with Fig.~4 shows that for the purely attractive potential
the significant pieces of the wave function are more shifted to
slightly larger momenta $p$ and $q$. This is consistent with the
larger binding energy given by the MT-IV potential, which causes
the 3N system to be tighter bound when viewed in configuration space
and thus giving a more extended structure in momentum space.
Due to the lack of repulsion this wave function does not change
sign.

\subsection{Momentum Distributions and Expectation Values}

In applications one does not access the wave function directly 
but rather only certain matrix elements thereof. One example, which 
can be `measured' at least approximately in electron scattering is the
momentum distribution, the probability to find a nucleon with 
momentum $q$ in the nucleus. In our case this is given by
\begin{eqnarray}
n(q) &=& 2 \pi q ^ 2 \ \int\limits_0^{\infty} dp \, p^2 \int\limits_{-1}^1 dx \,
                   \Psi^2(p,q,x)  \nonumber \\
    &=& \frac{1}{4\pi} \sum_l q ^ 2 \ \int\limits_0^{\infty} dp \, p^2 \,
                 \Psi^2_l(p,q,x). \label{eq:4.1a}
\end{eqnarray}
It is of interest to see how the sum of the partial wave form approaches
the expression evaluated directly from $\Psi(p,q,x)$. In Fig.~7 we
display the momentum distribution obtained from $\Psi(p,q,x)$
together with the partial wave sums corresponding to different $l$. The
small $q$ values are dominated by the $l = 0$ part. However, for a 
correct representation of $n(q)$ for larger $q$ larger $l$ values
are required. While for $q\leq 3$~fm$^{-1}$ $l = 0$ and 2 are sufficient,
at $q \sim 10$~fm$^{-1}$ one needs $l$ values as large as 8. 

Another interesting test for the numerical accuracy is the comparison of
a complementary quantity, namely
\begin{eqnarray}
\hat n(p) &=& 2 \pi \ p ^ 2 \ \int\limits_0^{\infty} dq \,q^2 \int\limits_{-1}^1 dx \,
  \Psi^2(p,q,x)  \nonumber \\
  &=& \frac{1}{4\pi}  \sum_l p ^ 2 \ \int\limits_0^{\infty} dq \, q^2 \,
     \Psi^2_l(p,q,x). \label{eq:4.1b}
\end{eqnarray}
This quantity does not seem to be easily accessible experimentally.
The momentum distribution $\hat n(p)$  as obtained from $\Psi(p,q,x)$
together with its partial wave sums  of different $l$ is shown in
Fig.~8. Here higher $l$ values are mostly needed to fill in the
dip at 2~fm$^{-1}$. 

A less detailed test for the quality of the 3N wave function is the
evaluation of the expectation value 
$\langle \Psi |H| \Psi \rangle \equiv \langle H \rangle$
and compare this value to the previously calculated eigenvalues.
Explicitly we evaluate the following expression
\begin{equation}
\langle \Psi |H| \Psi \rangle =  \langle \Psi |H_0| \Psi \rangle
      +    \langle \Psi | V | \Psi \rangle 
= 3 \,\langle \psi |H_0| \Psi \rangle
      +   3 \,\langle \Psi | V_1 | \Psi \rangle , \label{eq:4.2}
\end{equation}
where
\begin{equation}
\langle \psi |H_0| \Psi \rangle = 8 \pi^2 \int\limits_0^{\infty}
dp \int\limits_0^{\infty} dq \left [ \frac{1}{m}p^2 +\frac{3}{4m}q^2
\right] \int\limits_{-1}^1 dx \, p^2 \, q^2 \, \psi(p,q,x) \,
 \Psi(p,q,x) \label{eq:4.3}
\end{equation}
and
\begin{eqnarray}
\langle \Psi | V_1 | \Psi \rangle &=& 8 \pi^2 \int\limits_0^{\infty}
  dp \, p^2 \int\limits_0^{\infty} dq \, q^2 \int\limits_0^{\infty}
  dp' \, p'^2 \int\limits_{-1}^1 dx \int\limits_{-1}^1 dx' \nonumber
  \\ & & \times \, \Psi(p,q,x) \, v_1(p,p',x,x') \,
  \Psi(p',q,x'). \label{eq:4.4}
\end{eqnarray}
Here $v_1(p,p',x,x')$ is the expression for the potential containing
the integration over the azimuthal angle $\varphi$ (We use the same
notation as given in Ref.~\cite{3dt}),
\begin{equation}
v_1(q',q,x',x) = \int\limits_0^{\infty} d\varphi \;
V(q',q,x'x+\sqrt{1-x'^2} \, \sqrt{1-x^2}\cos \varphi).  \label{eq:4.5}
\end{equation}
In the case of the Malfliet-Tjon potential this integral can be performed
analytically.

The values of $\langle H \rangle$, $\langle H_0\rangle$, 
$\langle V\rangle$ 
are given in Table~V for both potentials MT-V and MT-IV calculated
in both schemes. One can see that the energy expectation values
and eigenvalues $E_{\rm t}$ agree with high accuracy within each
scheme as well as between the schemes.

\section{Verification of the  3N Schr\"odinger Equation}

In the previous sections we displayed properties of the 3N wave function,
which we obtained from the Faddeev component by evaluating
Eq.~(2.18). We showed that the expectation value $\langle H
\rangle$ evaluated with this wave function deviates from the calculated
eigenvalue by less than 0.1\%. However, since the expectation value
is obtained by integrating over
the wave function, this accuracy test gives information about
the overall quality of the wave function.
A more stringent test for the quality of $\Psi(p,q,x)$
at each point in the $p-q-x$ space is to determine the accuracy with
which $\Psi(p,q,x)$ fulfills the 3N Schr\"odinger equation. 
To the best of our knowledge, such a test
has never been carried out.

The 3N Schr\"odinger equation for the bound state of 3 identical
particles is given in Eq.~(\ref{eq:2.1}). Using the three different
sets of Jacobi momenta in momentum space we obtain
\begin{eqnarray}
E_{\rm t} \, \Psi(p_1,q_1,x_1)&=& \left[ \frac{1}{m}p_1^2
                            +\frac{3}{4m}q_1^2 \right]
                            \Psi(p_1,q_1,x_1)  \nonumber \\ && +
                            \int\limits_0^{\infty} dp'_1\, p_1^2
                            \int\limits_{-1}^1 dx'_1 \,
                            v_1(p_1,p'_1,x_1,x'_1) \,
                            \Psi(p'_1,q_1,x'_1)  \nonumber \\ && +
                            \int\limits_0^{\infty} dp'_2 \,p_2^2
                            \int\limits_{-1}^1 dx'_2 \,
                            v_2(p_2,p'_2,x_2,x'_2) \,
                            \Psi(p'_2,q_2,x'_2)  \nonumber \\ && +
                            \int\limits_0^{\infty} dp'_3 \,p_3^2
                            \int\limits_{-1}^1 dx'_3 \,
                            v_3(p_3,p'_3,x_3,x'_3) \,
                            \Psi(p'_3,q_3,x'_3). \label{eq:4.6}
\end{eqnarray}
Here $v_i(p_i,p'_i,x_i,x'_i), i=1,2,3$ contains the integration
over the azimuthal angle $\varphi$ in the different subsystems $i$ 
(Eq.~(\ref{eq:4.5})).
As an aside, in Ref.~\cite{kalusch} this equation was 
used to  obtain 
the eigenvalue $E_{\rm t}$ and the 3N wave function. We insert
 $\Psi(p,q,x)$ obtained from Eq.~(2.18) and verify the equivalence
of both sides of the Eq.~(5.1). For the numerical evaluation of the
integrals in Eq.~(5.1)  the $z$ direction 
for each integral is chosen separately, so that always 
${\bf q}_i$ points in the direction of $z$.
Then the integration vectors ${\bf p}'_i$ are chosen with respect
to ${\bf q}_i$ being the z-axis.
The momenta ${\bf p}_i$ and ${\bf q}_i$, $i=2,3$ have to be expressed
as functions of ${\bf p}_1$ and ${\bf q}_1$.
From the definitions of the momentum vectors in Eq.~(2.8) we obtain
\begin{eqnarray}
{\bf q}_2 &=& {\bf p}_1 -\textstyle{\frac{1}{2}} {\bf q}_1 
\nonumber  \\{\bf p}_2 &=& -\textstyle{\frac{1}{2}} {\bf p}_1
-\textstyle{\frac{3}{4}} {\bf q}_1  \nonumber \\ {\bf q}_3 &=& -{\bf
p}_1 + \textstyle{\frac{1}{2}}{\bf q}_1  \nonumber \\ {\bf p}_3 &=&
-\textstyle{\frac{1}{2}}{\bf p}_1 +\textstyle{\frac{3}{4}}{\bf q}_1.
\label{eq:4.7}
\end{eqnarray}
The corresponding angles are defined as 
$x_i = {\hat {\bf p}}_i \cdot {\hat {\bf q}}_i$, where $i = 2,3$. This
leads to 
\begin{eqnarray}
x_2&=& \frac{-p_1^2 +\frac{3}{4}q_1^2 -x_1p_1q_1 }
 {\sqrt{p_1^2 +\frac{9}{4}q_1^2 +3x_1p_1q_1} 
 \sqrt{p_1^2 +\frac{1}{4}q_1^2 -x_1p_1q_1}}  \nonumber \\
x_3&=&\frac{p_1^2 - \frac{3}{4}q_1^2 -x_1p_1q_1 }
{\sqrt{p_1^2 +\frac{9}{4}q_1^2 - 3x_1p_1q_1}
\sqrt{p_1^2 +\frac{1}{4}q_1^2 + x_1p_1q_1}}. \label{eq:4.8}
\end{eqnarray}

Since the functional form of the integrals in Eq.~(5.1)
is the same, we only evaluate the first, which we can view as
function of the momenta given in the coordinate system $1$
\begin{equation}
 \chi(p_1,q_1,x_1)=\int\limits_0^{\infty} dp'_1\, p_1^2
\int\limits_{-1}^1 dx'_1 \,v_1(p_1,p'_1,x_1,x'_1) \,\Psi(p'_1,q_1,x'_1).
\label{eq:4.9}
\end{equation}
The two remaining integrals in Eq.~(5.1) are then obtained by
interpolating on $\chi(p_i,q_i,x_i)$, $i=2,3$ using the coordinate
transformations given in Eqs.~(5.2) and (5.3). 

In Fig.~9 we display the relative error 
$\Delta=|(E\Psi(p,q,x)-H\Psi(p,q,x))| /| E\Psi(p,q,x)| \times 100$ 
for $x=-1$
for the wave function obtained from the MT-IV potential.  The graph
shows that the 3N Schr\"odinger equation is fulfilled with a 
numerical accuracy better than 1\% for momenta $p$ and $q$ up to
6~fm$^{-1}$. This is a typical range of momenta which enters 
calculations of matrix elements, e.g.,  for electron scattering. 
When $p$ exceeds values of 10~fm$^{-1}$ the relative error becomes
larger than 5\% for almost all values of $q$. This behavior can be
explained by inspecting the different Faddeev amplitudes separately
as they are given in Eq.~(2.19). If one considers the value $x=-1$
as shown in Fig.~9, one sees that for  $p=9.9$~fm$^{-1}$ and
$q=10.3$~fm$^{-1}$ the Faddeev amplitude $\psi_2$ is needed at
a value $q_2=15.05$~fm$^{-1}$, which is outside the $q$-range the
Faddeev equation was solved in this particular case. Instead of 
extrapolating, we set  $\psi_2(p,q,x)=0$, which introduces a small
discontinuity in the wave function. Though not visible in
integrations over the wave function,  this leads to a larger error when
considering the pointwise accuracy of the 3N Schr\"odinger equation.
Of course, this larger error in $\Psi(p,q,x)$ is completely irrelevant
in all practical cases, since for those large momenta $p$ and $q$
the wave function  
$\Psi(p,q,x)$ drops by about 8 orders of magnitude compared to its
value at the origin. In the relevant momentum regions the 3N
Schr\"odinger equation is fulfilled with very good accuracy by our
three dimensional wave function. This statement holds for all 
$x$ values.

For the above study we chose the wave function obtained from the
MT-IV potential, since this wave function does not have any node lines.
It is quite obvious that our definition of the relative error 
would give a large error at the locations where $\Psi(p,q,x)$
approaches zero. In fact, when calculating the relative error
with the wave function obtained from the MT-V potential, we can
identify the node lines of the wave function quite clearly.

\section{Summary}

An alternative approach to state-of-the-art 
three nucleon bound state calculations,
which are based on solving the Faddeev equations in a partial
wave truncated basis, is to work directly with momentum vector
variables.  We formulate the Faddeev equations for identical particles
as function of vector Jacobi momenta, specifically the magnitudes of the
momenta and the angle between them, and demonstrate their numerical
feasibility and the accuracy of their solutions. 
As two-body force we concentrated on a
superposition of an attractive and repulsive Yukawa interaction, which
is typical for nuclear physics, as well as on an attractive Yukawa
interaction. The corresponding two-body $t$-matrix, which enters the
Faddeev equations was also calculated as function of vector momenta. We
neglected spin degrees of freedom in all our calculations.

As first test for the numerical accuracy of the solution of the Faddeev
equation as function of vector variables, which is a three dimensional
integral equation, we determined the  energy eigenvalue of the bound
system and compared our result with the one obtained in a traditional
Faddeev calculation carried out in a partial wave truncated basis. We
achieved excellent agreement  between the two
approaches as well as excellent agreement with calculations in the
literature. We also found that the three dimensional Faddeev amplitude
is nearly independent of the angle between the two Jacobi momenta.
This is of course in agreement with the insights gained from the
approach in the partial wave scheme.

From the Faddeev amplitude we obtained the 3N wave function. We found
here also that the dependence on the angle between the two Jacobi
momenta is quite weak. We then performed a partial wave decomposition of
our three dimensional solutions for the Faddeev amplitude and the 3N
wave function and compared the so obtained partial wave amplitudes with
the ones directly calculated. Again we found excellent agreement between
the two approaches.  In order to further probe the quality of our wave
function we calculated the momentum distributions $n(q)$ and $\hat n(p)$ and
demonstrated that partial wave contributions up to $l=12$ are necessary
to build up the distribution $n(q)$ for large momentum transfers $q$.
In a similar vain we showed that higher partial waves are needed to 
fill in the sharp dip in $\hat n(p)$, which is obtained in a pure $s$-wave
calculation.

For a stringent test of the three dimensional wave function we inserted
it into the 3N Schr\"odinger equation and evaluated the accuracy with
which the eigenvalue equation is fulfilled throughout the entire
space where the solution is defined. We found that within the physical
relevant momentum region, namely $p$ and $q$ less than 10~fm$^{-1}$, the 
3N Schr\"odinger equation is fulfilled with high accuracy by our
numerical solution  of the Faddeev equation.

Summarizing we can state that the three dimensional Faddeev equation for
a bound state can be handled in a straightforward and numerically
reliable fashion. Once supplemented by spin degrees of freedom, this
approach will most likely be more easily implemented
 than the traditional partial
wave based method. State-of-the-art bound state calculations with
realistic nuclear force models typically require at least 34 channels.
The incorporation of 3N forces will most likely also be less cumbersome
in a three dimensional approach.

\vfill 
\acknowledgements 
This work was performed in part under the
auspices of the U.~S.  Department of Energy under contract
No. DE-FG02-93ER40756 with Ohio University, the NATO Collaborative
Research Grant 960892, the National Science Foundation under Grant
No. INT-9726624 and the Deutsche Forschungsgemeinschaft under Grant
GL-8727-1.  We thank the Ohio Supercomputer Center (OSC) for the use
of their facilities under Grant No.~PHS206, the National Energy
Research Supercomputer Center (NERSC) for the use of their facilities
under the FY1998 Massively Parallel Processing Access Program and the
H\"ochstleistungsrechenzentum J\"ulich for the use of their Cray-T3E.
The authors would like to thank H.~Kamada for many stimulating and
fruitful discussions.

%----------------------------------------------------

\appendix
\section{Explicit Representation of the partial wave projected
Permutation Operator}

For the convenience of the reader we provide the quantities $G$ and
$\tilde G$ related to the permutation operators for general total
angular momentum $L$.

The function $G_{ll'}(q,q',x')$ from Eq.~(\ref{partialfad}) is a
combination of Legendre polynomials $P_k(x)$:
\begin{eqnarray}
G_{l\lambda,l'\lambda'} (q,q',x) = \sum_k P_k(x) \sum_ {\mu_1+\mu_2=l} \sum_
{\nu_1+\nu_2=l'} \ q'^{\mu_1+\nu_2} \ q^{\mu_2+\nu_1} \
g_{l\lambda,l'\lambda'}^{k \mu_1 \nu_1  \mu_2 \nu_2}.
\end{eqnarray}

For general $L$ the geometrical coefficient is given as 
\begin{eqnarray}
g_{l\lambda,l'\lambda'}^{k \mu_1 \nu_1 \mu_2 \nu_2} & = & \sum_{gg'} \hat k \, 
     \sqrt{ \hat l \hat \lambda \hat l ' \hat \lambda '} \,
     \sqrt{ \frac{\hat l ! \hat l ' !}{(2 \mu_1)! \, (2 \mu_2)! \,
           (2 \nu_1)! \, (2 \nu_2)! } } \nonumber \\[5pt]
& & \times \, (-)^{l'} \left( {\textstyle{\frac{1}{2}}} 
     \right)^{\mu_2+\nu_2}
     \left\{ \begin{array}{ccc} \mu_1   & \mu_2 & l \cr
                                \lambda &  L    & g \end{array}
     \right\} \ 
     \left\{ \begin{array}{ccc} \nu_1   & \nu_2 & l' \cr
                                \lambda'&  L   & g' \end{array}
     \right\} \ 
     \left\{ \begin{array}{ccc} \mu_1   & g & L \cr
                                \nu_1   & g'& k  \end{array}
     \right\} \ \nonumber \\[5pt]
& &  \times \, C( \mu_2 \ \lambda \ g   , 0 \ 0)  \
     C( \nu_2 \ \lambda' \ g' , 0 \ 0)  \ 
     C( k \ \nu_1 \ g   , 0 \ 0)  \
     C( k \ \mu_1 \ g'  , 0 \ 0).
\end{eqnarray}

\noindent
Here $l$ and $\lambda$ are the relative orbital angular momenta
related to $p$ and $q$. We also use the notation $\hat l \equiv 2l+1$.

In our context we have $L = 0$, which leads to $l=\lambda$. Then
$g$ reduces to
\begin{eqnarray} 
g_{ll'}^{k \mu_1 \nu_1  \mu_2 \nu_2} & = &  
      \sqrt{\frac{\hat l \hat l ' }{\hat \mu_1 \ \hat \nu_1 }}  \  
     \sqrt{ \frac{\hat l ! \hat l ' !}{(2 \mu_1)! \, (2 \mu_2)! \,
           (2 \nu_1)! \, (2 \nu_2)! } } \nonumber \\[5pt]
& & \times \, \left( {\textstyle{\frac{1}{2}}} \right)^{\mu_2+\nu_2} \ (-)
    ^ { \mu_1 + \nu_1}      
     C( \mu_2 \ l  \ \mu_1   , 0 \ 0)  \
     C( \nu_2 \ l' \ \nu_1   , 0 \ 0)  \ 
     C( \mu_1 \ \nu_1 \ k  , 0 \ 0) ^ 2   \
\end{eqnarray}

\noindent
and $G_{l\lambda,l'\lambda'}$ to $G_{l l'}$.

The quantity $\tilde G(p,q,x)$ occurs in Eq.~(\ref{partialwave}).
In the case of general $L$ the function $\tilde G$ reads
\begin{eqnarray}
\tilde G_{l\lambda,l'\lambda'} (p,q,x)  = \sum_k P_k(x) \sum_ {\mu_1+\mu_2=l'} \sum_
{\nu_1+\nu_2=\lambda '} \ p^{\mu_1+\nu_1} \ q^{\mu_2+\nu_2} \
\tilde g_{l\lambda,l'\lambda'}^{k \mu_1 \nu_1  \mu_2 \nu_2}
\end{eqnarray}
with the geometrical factor
\begin{eqnarray}
\tilde g_{l\lambda,l'\lambda'}^{k \mu_1 \nu_1  \mu_2 \nu_2} & = & \sum_{gg'} 
     \hat k  \,
     \sqrt{ \hat l '  \hat \lambda ' \hat g \hat g '}
     \sqrt{ \frac{\hat l ' ! \hat \lambda ' !}{(2 \mu_1)! \, (2 \mu_2)!  \,
           (2 \nu_1)! \, (2 \nu_2)! } } \nonumber \\[5pt] 
& &  \times \, (-)^{g+L+\nu_2} 
     \left( {\textstyle{\frac{1}{2}}} \right)^{\mu_1+\nu_2}
     \left( {\textstyle{\frac{3}{4}}} \right)^{\mu_2}
     \left\{ \begin{array}{ccc} \mu_1   & \mu_2 & l ' \cr
                                \nu_1   & \nu_2 & \lambda ' \cr
                                   g    & g'    & L     \end{array} 
     \right\} \ 
     \left\{ \begin{array}{ccc} g'    &    g    & L  \cr
                                l    & \lambda & k  \end{array}
     \right\} \ \nonumber \\[5pt]
& &  \times \, C( \mu_1 \ \nu_1   \ g       , 0 \ 0)  \
     C( \mu_2 \ \nu_2   \ g'      , 0 \ 0)  \
     C( g     \ k       \ l       , 0 \ 0)  \
     C( g'    \ k       \ \lambda , 0 \ 0).
\end{eqnarray}

\noindent
Again, for $L = 0$ $\tilde g$ reduces to
\begin{eqnarray}
\tilde g_{l l'}^{k \mu_1 \nu_1 \mu_2 \nu_2 } & = & \sum_{g} \hat k   \,
     (-) ^ { \mu_2+g+k} \,
     \left( {\textstyle{\frac{1}{2}}} \right)^{\mu_1+\nu_2} \,
     \sqrt{\frac{\hat l'}{\hat l }} 
     \sqrt{ \frac{\hat l ' ! \hat \lambda ' !}{(2 \mu_1)! \, (2 \mu_2)!  \,
           (2 \nu_1)!\, (2 \nu_2)! } } \nonumber \\[5pt] 
& & \times \, 
     \left( {\textstyle{\frac{3}{4}}} \right)^{\mu_2}
     \left\{ \begin{array}{ccc} \mu_1   & \mu_2 & l ' \cr
                                \nu_2   & \nu_1 & g   \end{array} 
     \right\} \ 
     C( \mu_1 \ \nu_1   \ g       , 0 \ 0)  \
     C( \mu_2 \ \nu_2   \ g       , 0 \ 0)  \
     C( g     \ k       \ l       , 0 \ 0)^2  
\end{eqnarray}

\noindent
and correspondingly $\tilde G_{l\lambda,l'\lambda'}$ to $\tilde G_{l l'}$.

%----------------------------------------------------

%\pagebreak

\pagebreak
%%%%%%%%%%%%%%%%%%%%%%%%%%%%%%%%%%%%%%%%%%%%%%%%%%%%%%

\begin{table}
\caption{ Parameters of the Malfliet-Tjon type potentials. As conversion
factor we use units such that $\hbar c = 197.3$ MeV fm = 1. We also use
$\hbar^2/m = 41.470$~MeV fm$^2$.}

\begin{tabular}{l|cccc}
\mbox{ }&$V_{\rm A}$ [MeV fm]&$\mu_{\rm A}$ [fm$^{-1}$]&$V_{\rm R}$
 [MeV fm] & $\mu_{\rm R}$ [fm$^{-1}$] \\ \hline MT-V &
     -570.3316 & 1.550 & 1438.4812 & 3.11 \\ MT-IV &
     \phantom{0}-65.1090 & 0.633 & - & - \\
\end{tabular}
\end{table}

\begin{table}
\caption{The calculated eigenvalue $E_{\rm t}$ of the Faddeev equation
as function of the number of grid points NP1, NP2, NQ1, NQ2, and NX
chosen for the solution of Eq.~(\ref{eq:2.13}).  The calculations are
based on the potentials MT-V and MT-IV.}
\begin{tabular} {ccccc|cc}
NP1 & NP2 & NQ1 & NQ2 & NX & $E_{\rm t}$ [MeV] (MT-V) & $E_{\rm t}$
[MeV] (MT-IV) \\
\hline
20 &12 &  20 &12 &40 &  -7.74387 & -25.0416 \\
24 &24 &  24 &16 &40 &  -7.74000 & -25.0453 \\
32 &24 &  24 &16 &40 &  -7.73761 & -25.0485 \\ 
32 &24 &  32 &16 &40 &  -7.73761 & -25.0483 \\
32 &24 &  32 &32 &40 &  -7.73761 & -25.0483 \\
32 &32 &  32 &32 &40 &  -7.73761 & -25.0483 \\
48 &32 &  48 &32 &40 &  -7.73666 & -25.0499 \\
64 &32 &  48 &32 &40 &  -7.73666 & -25.0502 \\
64 &32 &  64 &32 &40 &  -7.73650 & -25.0502 \\
72 &36 &  64 &32 &40 &  -7.73650 & -25.0499 \\
\end{tabular}
\end{table}
 
\begin{table}
\caption{The calculated eigenvalue $E_{\rm t}$ of the Faddeev equation
as function of the number of partial waves employed. A fixed set of
grid points is chosen corresponding to NP1~=~64, NP2~=~32, NQ1~=~64,
NQ2~=~32, NX~=~16. The partial wave $t$-matrix is calculated on a
larger grid of $160 \times 160$ grid points. The calculations are based
on the potentials MT-V and MT-IV.}
\begin{tabular} {c|c|c}
$l $ &  $E_{\rm t}$ [MeV] (MT-V) & $E_{\rm t}$ [MeV] (MT-IV)\\
\hline
\phantom{1}0& -7.53975   & -24.8616 \\
\phantom{1}2& -7.71470   & -25.0465 \\ 		
\phantom{1}4& -7.73383   & -25.0552 \\		
\phantom{1}6& -7.73613   & -25.0562 \\		
\phantom{1}8& -7.73649   & -25.0564 \\		
          10& -7.73656   & -25.0565 \\		
          12& -7.73658   & -25.0565 \\
\end{tabular}
\end{table}

\begin{table}
\caption{The relative contributions of the partial wave Faddeev
amplitudes $F_l$ for each partial wave. We give also the corresponding
values $W_l$ for the wave function.  The calculations are based on
the potentials MT-V and MT-IV.}
\begin{tabular} {c|l|l|l|l}
     & \multicolumn{2}{c|}{MT-V} & \multicolumn{2}{c}{MT-IV} \\
$l $ & $ F_l$ & $ W_l$ & $ F_l$ & $ W_l$  \\
\hline
\phantom{1}0& $99.9951$                 & $99.0851$ & $99.9940$
& $99.2121$ \\
\phantom{1}2& $0.4823 \cdot  10 ^{-2}$  & $0.7482$ &  $0.5923 \cdot  10 ^{-2}$
& $0.7318$  \\ 
\phantom{1}4&  $0.7988 \cdot  10 ^{-4}$ & $0.1159$ &  $0.3128 \cdot  10 ^{-4}$
& $0.4682\cdot  10 ^{-1}$\\
\phantom{1}6&  $0.2278 \cdot  10 ^{-5}$ & $0.3305 \cdot  10 ^{-1}$ &  $0.8813
 \cdot  10 ^{-6}$
& $0.6743\cdot  10 ^{-2}$ \\
\phantom{1}8&  $0.1110 \cdot  10 ^{-6}$ & $0.1088 \cdot  10 ^{-1}$ &  $0.6056 
\cdot  10 ^{-7}$
& $0.1520\cdot  10 ^{-2}$ \\
          10&  $0.8677 \cdot  10 ^{-8}$ & $0.3939 \cdot  10 ^{-2}$ &  $0.7156 
\cdot  10 ^{-8}$
& $0.4528\cdot  10 ^{-3}$ \\
          12&  $0.1006 \cdot  10 ^{-8}$ & $0.1545 \cdot  10 ^{-2}$ &  $0.1147 
\cdot  10 ^{-8}$
& $0.1627\cdot  10 ^{-3}$\\
\end{tabular}
\end{table}

\begin{table}
\caption{The expectation value $\langle H\rangle$, $\langle H_0\rangle$,
and $\langle V\rangle$ calculated for the potentials MT-V and MT-IV
within the three dimensional scheme (3D) and the partial wave scheme
(PW).}
\begin{tabular} {cc|cccc}
 & & $\langle H_0\rangle$~[MeV] & $\langle V\rangle$~[MeV] & 
 $\langle H\rangle$~[MeV] & $(E_{\rm t}-\langle H\rangle)$~[MeV]\\
  \hline
MT-V& 3D & 29.77706   &  -37.51340  &  -7.73634  & 0.00011 \\
    & PW & 29.7776    &  -37.5139   &  -7.73634  & 0.00024 \\
\hline \hline
MT-IV & 3D & 77.2055 & -102.2550 & -25.0495 & 0.0003    \\
      & PW & 77.2653 & -102.3207 & -25.0554 & 0.0011    \\
\end{tabular}
\end{table}

\pagebreak
%%%%%%%%%%%%%%%%%%%%%%%%%%%%%%%%%%%%%%%%%%%%%%%%%%%%%%

\noindent
\begin{figure}
\caption {The angular dependence for the symmetrized off-shell
 $t$-matrix, $t_{\rm s}(p,p',x,E)$, is displayed 
for $p=0.75$~fm$^{-1}$ and  $E=-50$ MeV.  \label{fig1}}
\end{figure}

\noindent
\begin{figure}
\caption {The magnitude of the
Faddeev component $\psi(p,q,x)$ for $x=1$ calculated
from the MT-V potential. \label{fig2}}
\end{figure}

\noindent
\begin{figure}
\caption {The partial wave projected Faddeev components 
$\psi_l(p,q)$ calculated from the MT-V potential are shown
for $l=0,2,4$. The
left column of graphs represents the amplitudes obtained
from the three dimensional calculations. Here (a) stands for
the $l=0$ amplitude, (b) the one for $l=2$, and (c) the one for $l=4$.
The amplitudes
in the right column are obtained as solutions of the partial
wave Faddeev equation, (d) represent the $l=0$ amplitude,
(e) the one for $l=2$, and (f) the one for $l=4$.   \label{fig3}}
\end{figure}

\noindent
\begin{figure}
\caption {The magnitude of the
3N bound state wave function $\Psi(p,q,x)$ for $x=1$
calculated from the MT-V potential.\label{fig4}}
\end{figure}

\noindent
\begin{figure}
\caption {The partial wave projected bound state wave functions
$\Psi_l(p,q)$ calculated from the MT-V potential are show
for $l = 0,2,4$.
The left column of graphs represents the amplitudes obtained
from the three dimensional calculations. Here (a) stands for
the $l=0$ amplitude, (b) the one for $l=2$, and (c) the one for $l=4$.
The amplitudes
in the right column are obtained as solutions of the partial
wave Faddeev equation, (d) represent the $l=0$ amplitude,
(e) the one for $l=2$, and (f) the one for $l=4$.
 \label{fig5}}
\end{figure}

\noindent
\begin{figure}
\caption {The magnitude of the
3N bound state wave function $\Psi(p,q,x)$ for $x=1$
calculated from the MT-IV potential.\label{fig6}}
\end{figure}

\noindent
\begin{figure}
\caption {The momentum distribution $n(q)$ is shown in (a) for
the MT-V potential as calculated in the two different schemes.
The solid line refers shows the result from the three dimensional
calculation whereas the dashed line gives the result from the
partial wave sum up to $l=12$. The monotonic increasing partial wave sums 
for $l=0,2,4,6,8,10$, and $12$ are shown in (b). \label{fig7}}
\end{figure}

\noindent
\begin{figure}
\caption {The momentum distribution $\hat n(p)$ is shown in (a) for
the MT-V potential as calculated in the two different schemes.
The solid line refers shows the result from the three dimensional
calculation whereas the dashed line gives the result from the
partial wave sum up to $l=12$. The monotonic increasing partial wave sums 
for $l=0,2,4,6,8,10$, and $12$ are shown in (b). \label{fig8}}
\end{figure}

\noindent
\begin{figure}
\caption {The relative error $\Delta$  as defined in the text 
as function of the momenta $p$ and $q$ for the fixed angle $x=-1$ 
represents a measure for 
the accuracy with which the 3N Schr\"odinger equation is fulfilled
by the three dimensional wave function obtained from the MT-IV
potential. The numbers at the contour lines give the relative
error in \%. \label{fig9}}

\end{figure}

%\end{document}

\input{psfig}

\pagebreak

\psfig{file=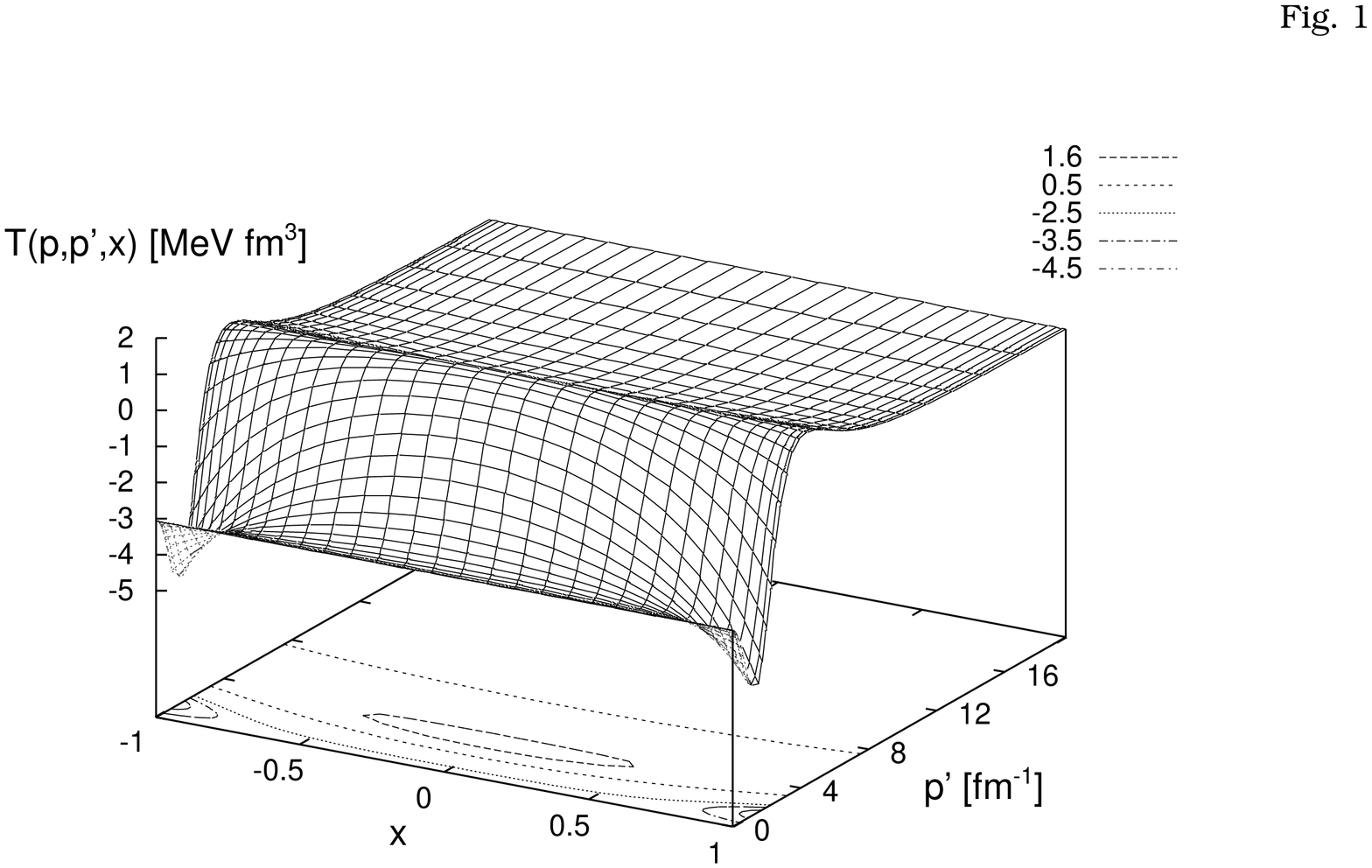,width=300pt,angle=-90}
\vspace{-1.5cm}
\psfig{file=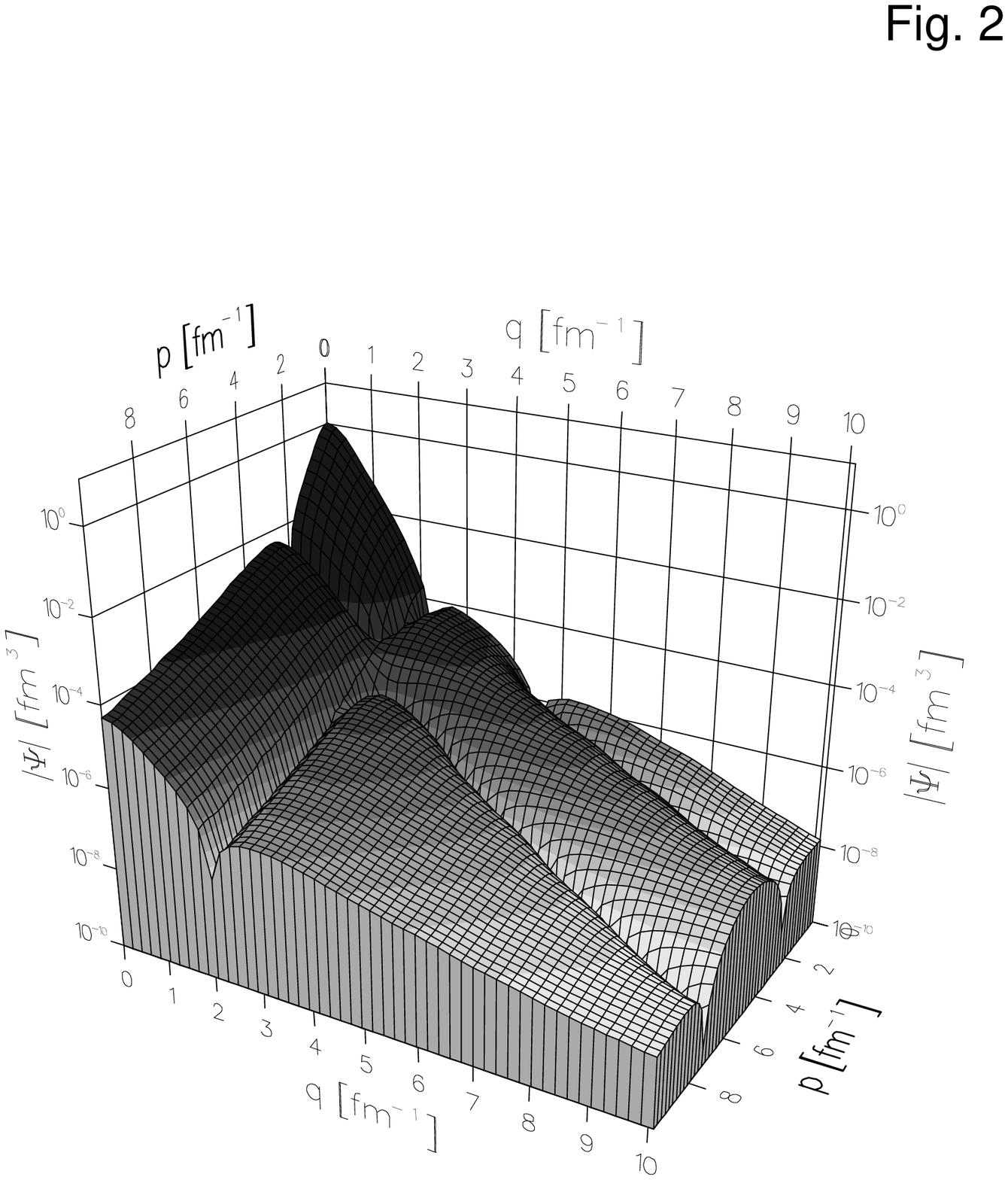,width=300pt}
\vspace{-9.5cm}

\pagebreak

\psfig{file=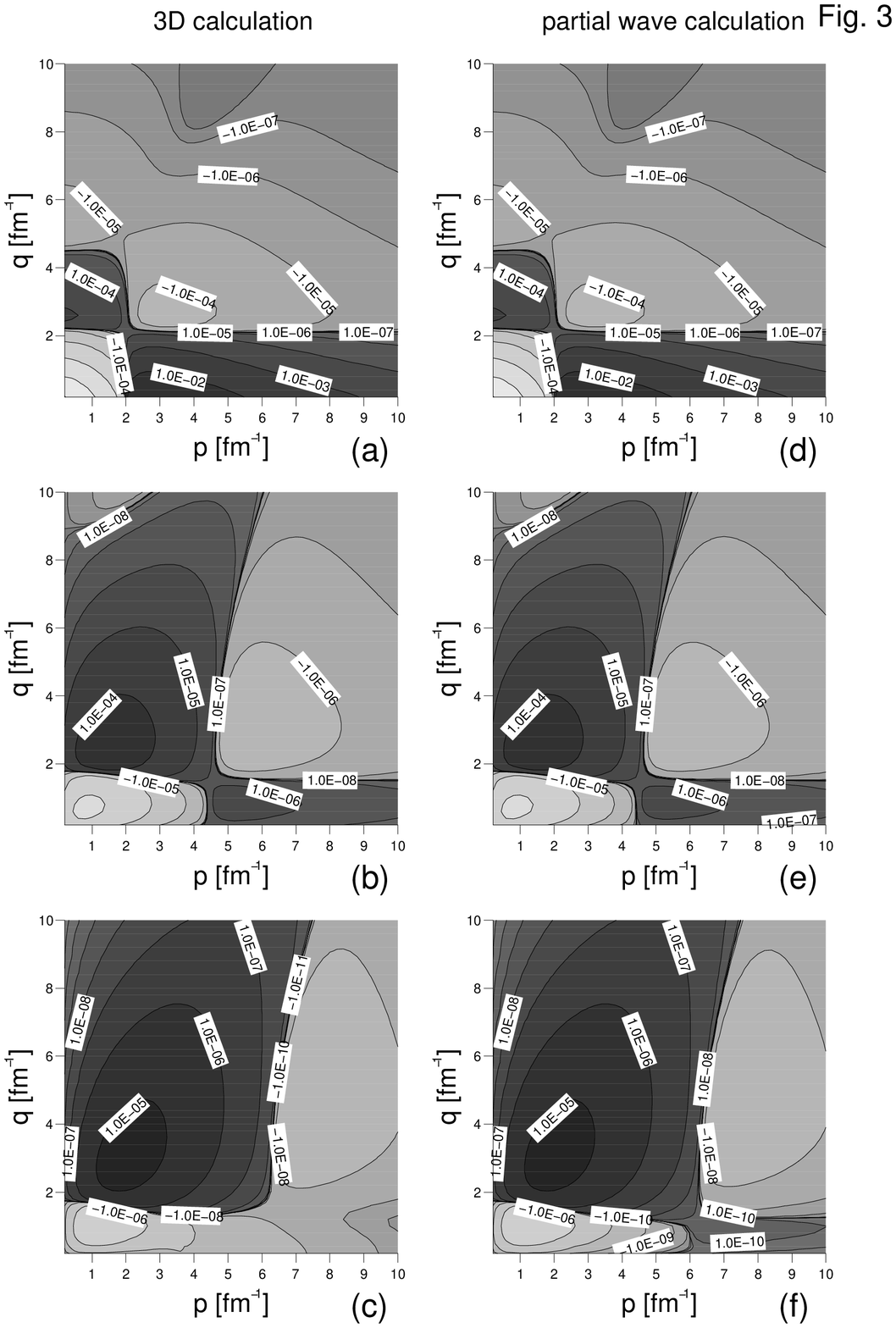,width=400pt}

\pagebreak
\psfig{file=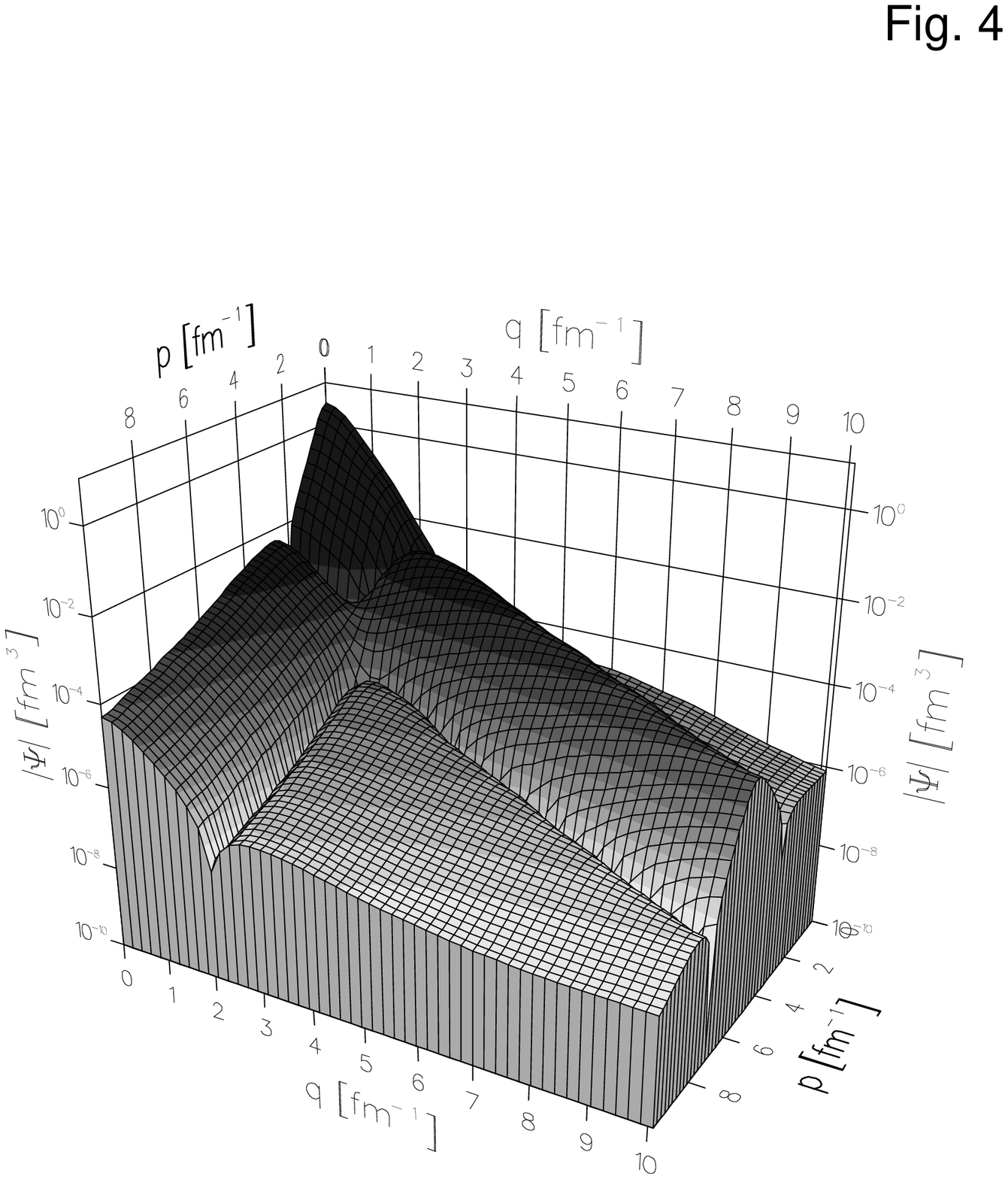,width=300pt}

\pagebreak

\psfig{file=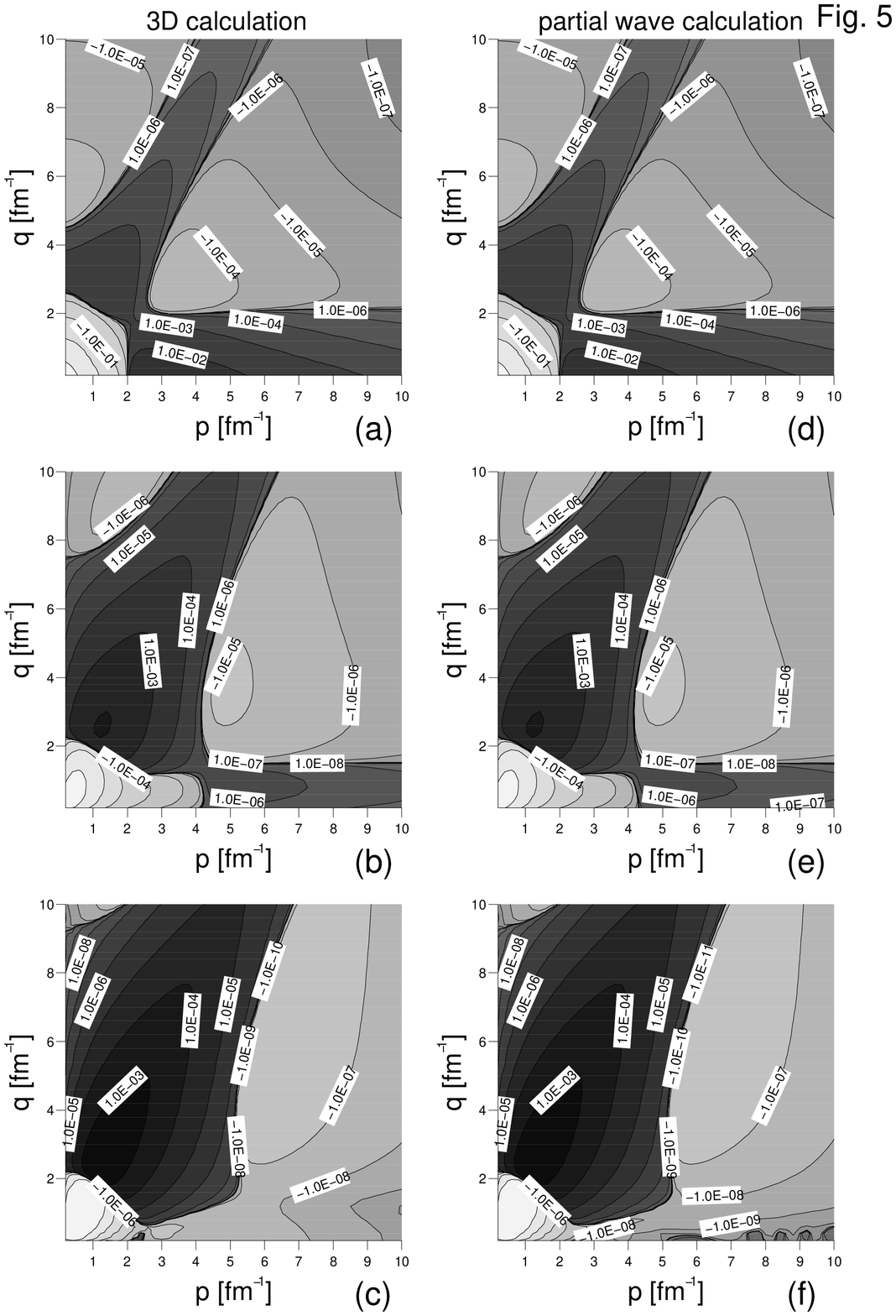,width=400pt}

\pagebreak
\psfig{file=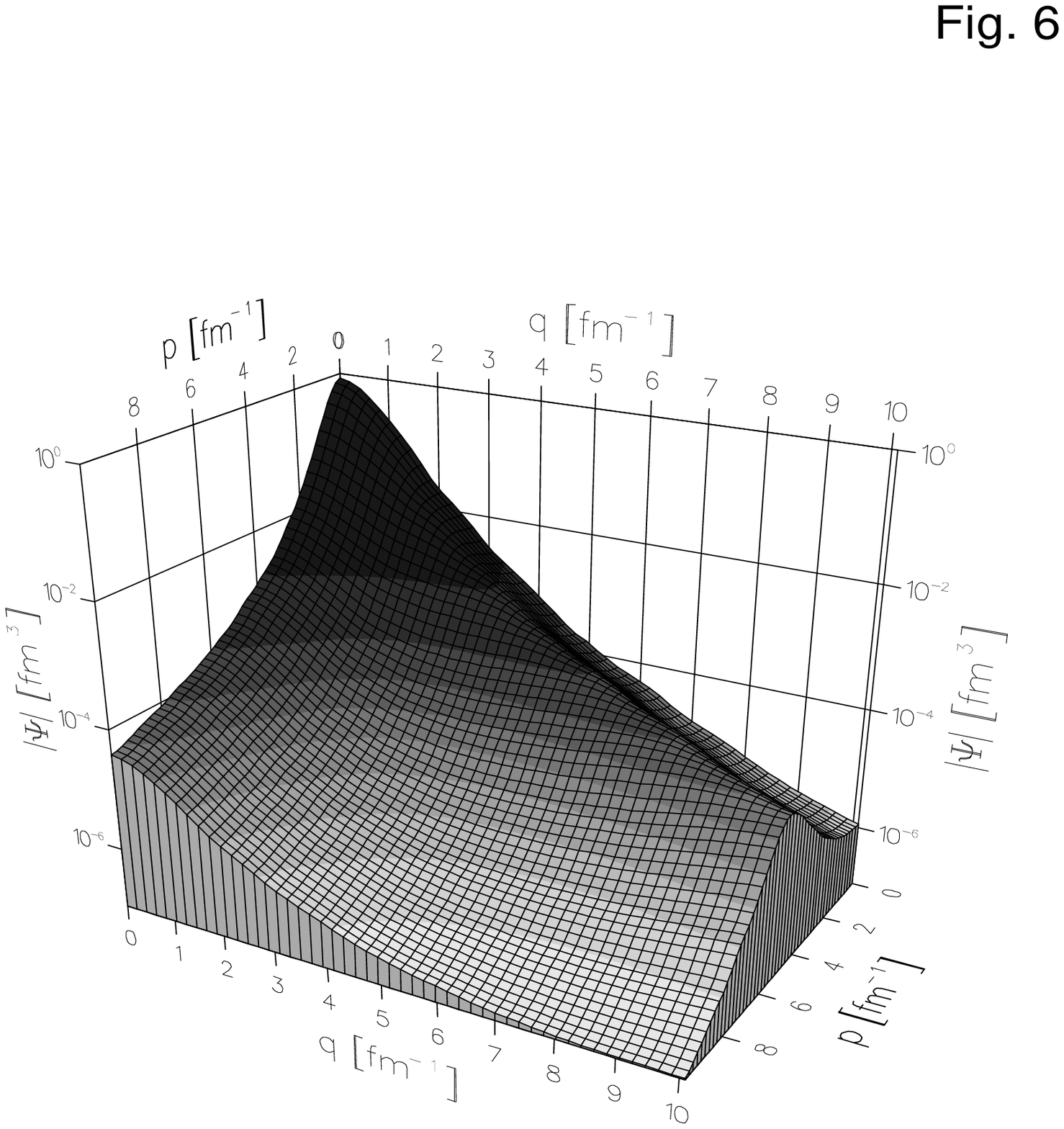,width=300pt}
\vspace{-7.5cm}
\psfig{file=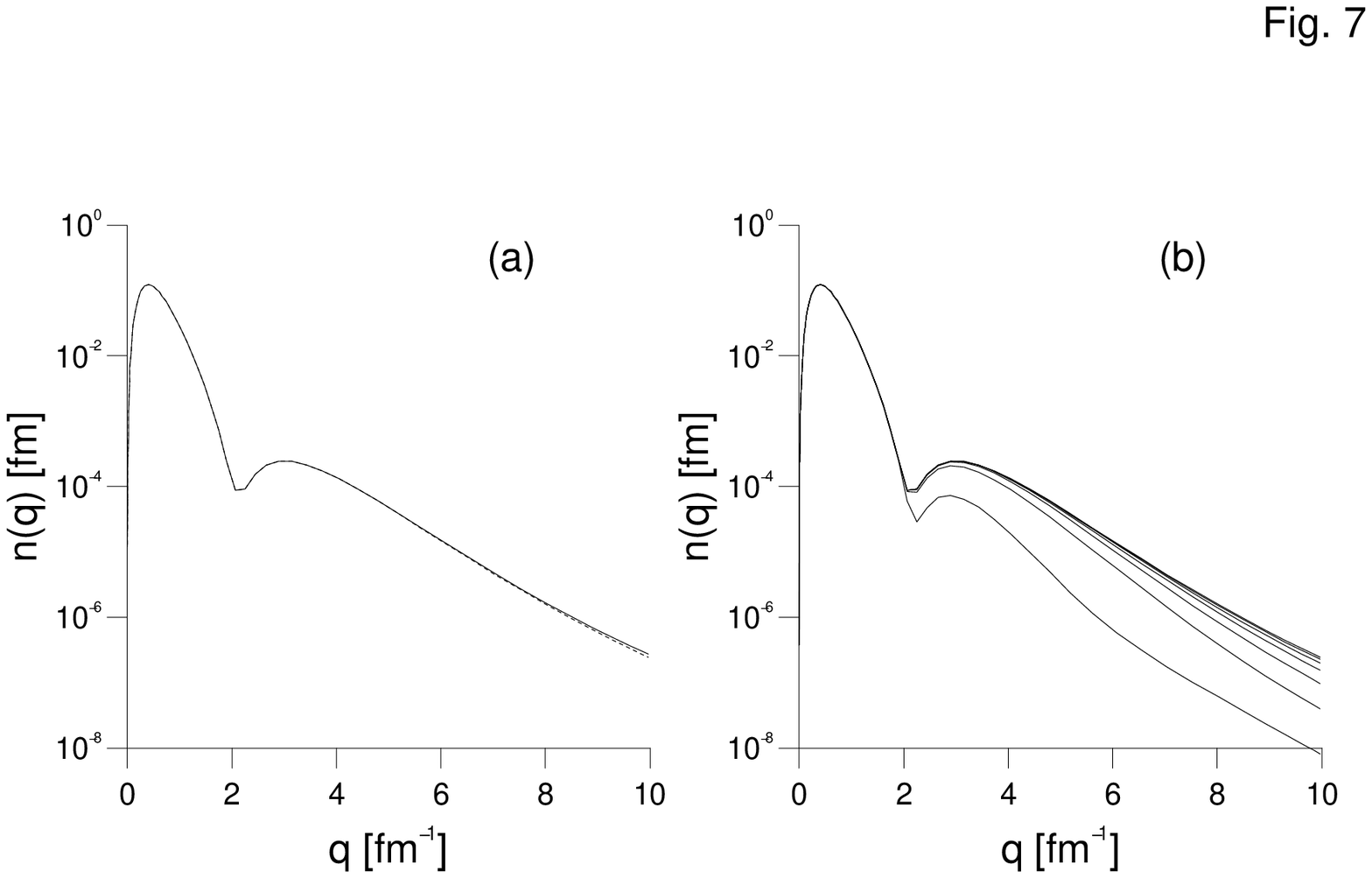,width=300pt}
\pagebreak
\phantom{0}
\vspace{-7.5cm}
\psfig{file=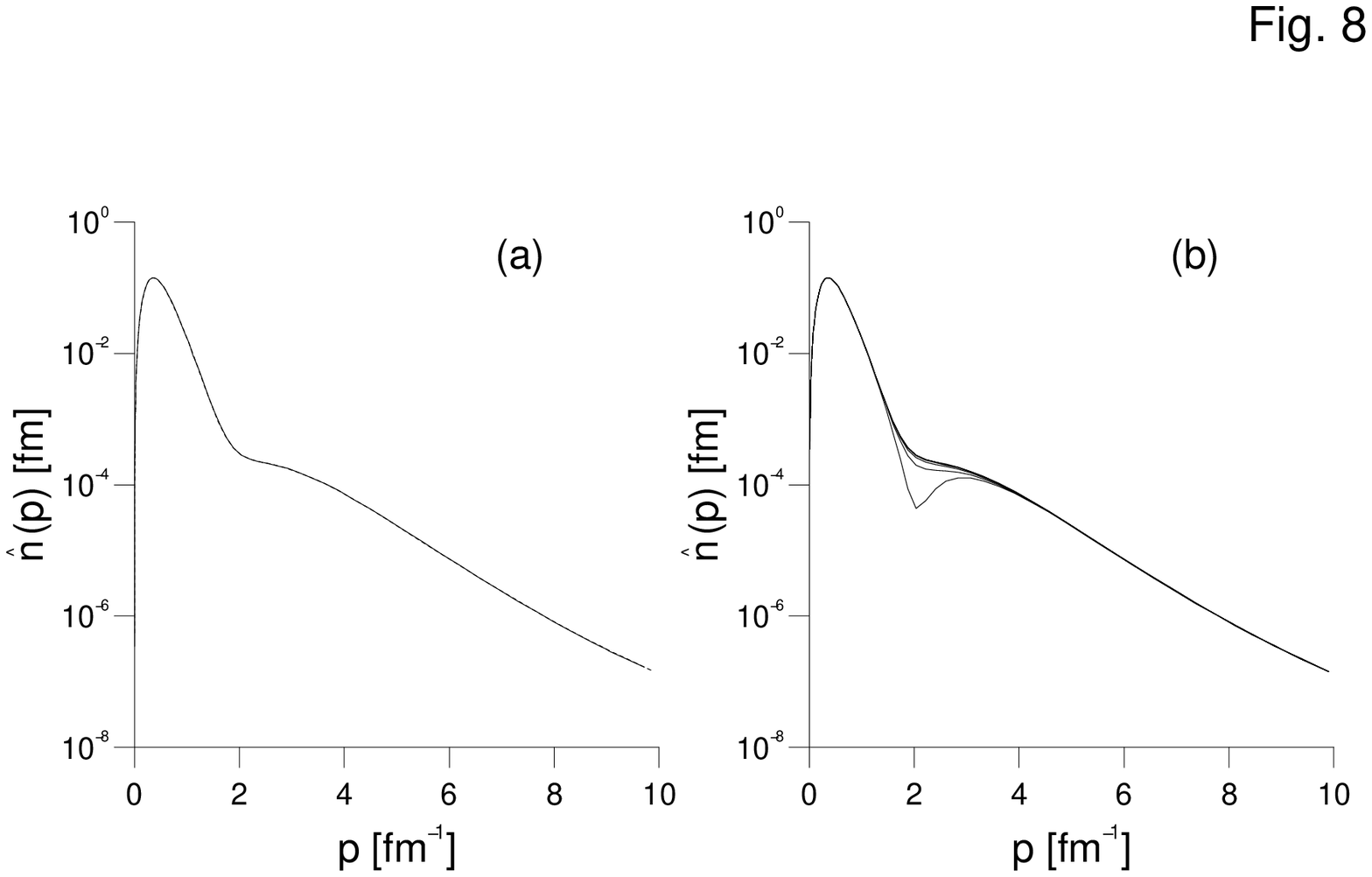,width=300pt}
\vspace{-4.5cm}
\psfig{file=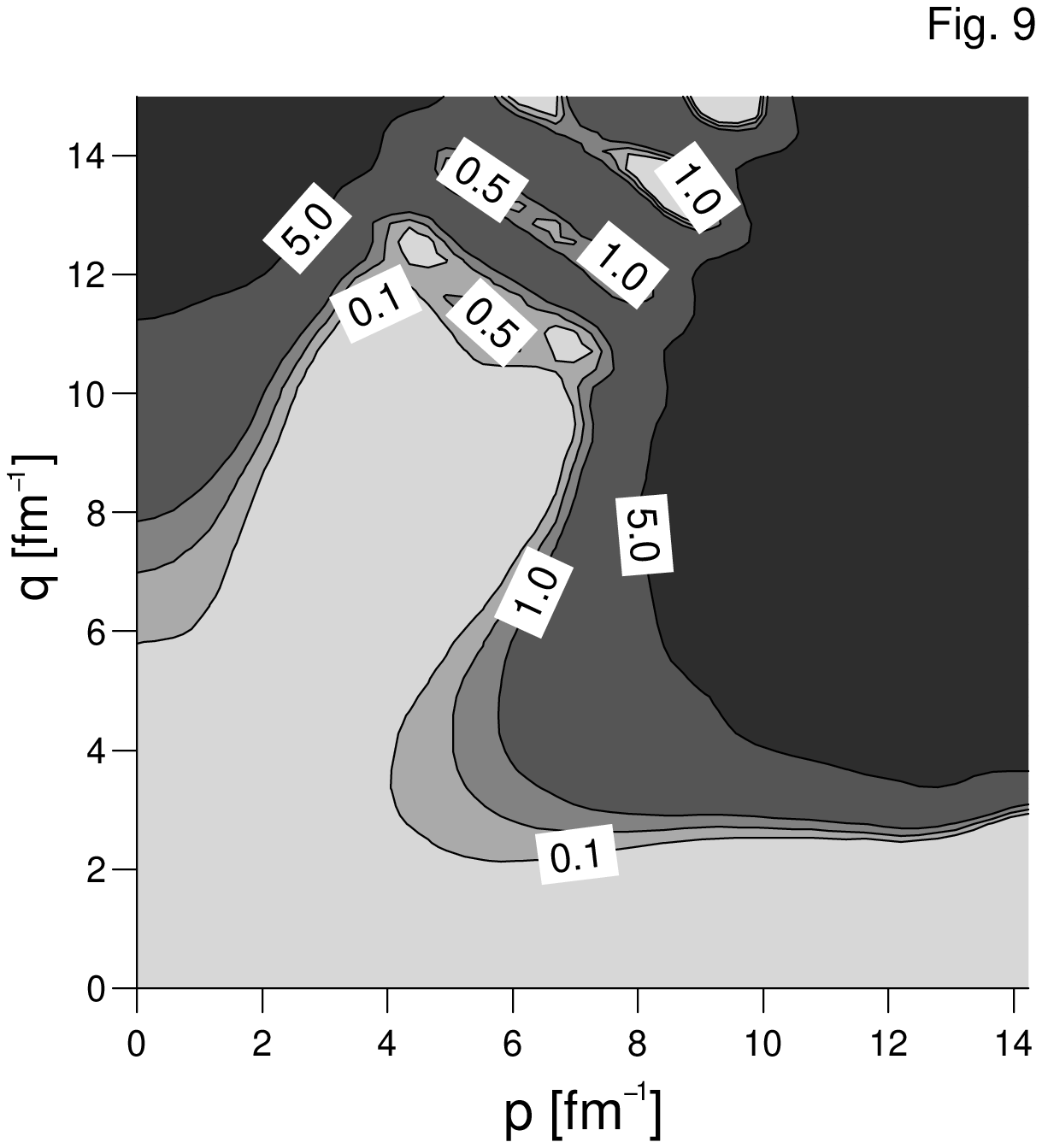,width=300pt}
\vspace{-7.5cm}

\end{document}